\documentstyle[preprint,aps,prb]{revtex}

\begin{document}

\newcommand{\beq}{\begin{equation}}
\newcommand{\eeq}{\end{equation}}
\draft

\title
{A microscopic approach to the response of  $^{\bf 3}$He-$^{\bf 4}$He
mixtures}
\author{A. Fabrocini and L.Vichi}
\address{
Department of Physics, University of Pisa,
INFN Sezione di Pisa, I-56100 Pisa, Italy}
\author{ F.Mazzanti and A. Polls}
\address{
Departament d'Estructura i Constituents de la Mat\`eria, Universitat de
Barcelona, \protect\\
Diagonal 647, E-08028 Barcelona, Spain}


\maketitle

\begin{abstract}
Correlated Basis Function perturbation theory is used to
evaluate the zero temperature response $S(q,\omega)$ of $^3$He-$^4$He
mixtures for inelastic neutron scattering, at  momentum transfers
$q$ ranging from $1.1$ to $1.7 \AA^{-1}$. We adopt a Jastrow correlated
ground state and a basis of correlated particle-hole and phonon states.
We insert correlated one particle-one hole and one-phonon states to
compute the second order response. The decay of the one-phonon states into
two-phonon states is accounted for in boson-boson approximation.
The full response is splitted into three partial components
$S_{\alpha \beta}(q,\omega)$, each of them
showing a particle-hole bump and a one phonon, delta shaped peak, which
stays separated from the multiphonon background. The cross term
$S_{34}(q,\omega)$ results to be of comparable importance to
$S_{33}(q,\omega)$ in the particle-hole sector and to $S_{44}(q,\omega)$
in the phonon one. Once the one-phonon peak has been
convoluted with the experimental broadening, the computed scattering
function is in semiquantitative agreement with recent experimental
measurements.
\end{abstract}



\narrowtext

{\bf 1. INTRODUCTION}

Isotopic atomic Helium mixtures are an intriguing case for many-body
physicists. There exists a large body of experimental data, concerning
mostly static properties (for instance, the chemical potentials and
the maximum solubility \cite{Ebner70,Baym78,DeBruyn87}).
Excitation spectra and related quantities,
as the zero concentration ($x_3=0$) $^3$He effective mass ($m_3^*$),
have been also measured \cite{Hilton77}. Recently, inelastic neutron scattering
experiments have been carried out  both at low, or intermediate,
and high momentum transfers \cite{Fak90,Sokol94}.  In both regions the
measured response presents two generally distinguishable structures,
to be ascribed to boson like collective excitations (phonons and rotons)
and to Fermi particle-hole ones. However, this apparently simple picture
hides a large interplay between the components of the mixtures, each of
them probably contributing on comparable foot to both the branches of the
response. The reason for this lies in the large correlation effects, which
are present in the system because of the strong interatomic potential and
of the large density. This are also the motivations why truly microscopic and
ab initio studies of Helium mixtures are difficult, and, in the case of
the response, practically absent in literature.

Qualitative studies of the response have been done in ref.\cite{Szprynger85}
(using a matrix dispersion-relation representation)and in
ref.\cite{Saarela_rep}
with a correlated RPA approach (very similar, in spirit, to the
phenomenological Polarization Potential method used in ref.\cite{Hsu}.
Here we will employ the Correlated Basis Function
(CBF) perturbation theory, to embody the above correlation effects directly
into the basis functions. CBF has shown to be a powerful tool to
succesfully study Helium at zero temperature: the energetics of both pure
$^4$He and $^3$He are well described by sophisticated correlated ground state
wave functions, containing explicit two--, three--body, back--flow and
spin correlations \cite{FantoniHe4,FantoniHe3,VivianiHe3}; properties
of one $^3$He impurity in $^4$He, such as chemical potential and effective
mass are also quantitatively reproduced by such correlated wave functions
\cite{Fabrocini86}. In particular, by using CBF based perturbation theory,
with the insertion of up to two correlated independent phonon intermediate
states, the impurity effective mass $m_3^*$ turns out to be $2.2 m_3$, to
be compared with the experimentally measured $2.3 m_3$ value.

The behavior of the $^3$He effective mass with the concentration
in dilute mixtures
has been recently object of some debate. Specific heat measurements
\cite{Greywall79,Owers88} at finite $x_3$ do not show appreciable
deviations from its $x_3=0$ value. In ref.\cite{Fak90} the Authors have
to postulate a much larger value ($m_3^*\sim 2.95 m_3$ at $x_3=0.05$) in
order to reproduce the position of the particle--hole response with
a Lindhard like function and using a simple Landau--Pomeranchuk (LP)
quasiparticle spectrum \cite{Landau48},

\begin{equation}
\epsilon_k(LP)=\, \epsilon_0 \, + \, \frac {\hbar^2 k^2}{2 m_3^*} \, .
\label{eq:LP}
\end {equation}

 This contradiction does not appear if one modifies the LP spectrum
( LP modified, or LPM) as:

\begin{equation}
\epsilon_k(LPM)=\, \epsilon_0 \, + \, \frac {\hbar^2 k^2}{2 m_3^*}\,
\frac{1}{ 1\,+\, \gamma\ k^2} .
\label{eq:LPM}
\end {equation}

 There are both experimental \cite{Fak90,Hilton77} and theoretical
 \cite{Bhatt78} indications of a deviation from the simple LP
 form.

 In a CBF based approach, we assume to have an homogeneous mixture of
 $N_3$ $^3$He atoms and $N_4$ $^4$He atoms in a volume $\Omega$, with partial
 densities $\rho_{\alpha=3,4}=N_\alpha /\Omega$, total density
 $\rho=\rho_3+\rho_4$ and concentrations $x_\alpha=\rho_\alpha /\rho$.
 We will keep constant densities, while letting $N_\alpha$ and
 the volume going to infinity. The nonrelativistic Hamiltonian of the
 mixture is

\begin {equation}
H=-\sum_{\alpha =3,4} \sum_{i=1}^{N_\alpha}
\frac {\hbar^2}{2 m_\alpha} \nabla_{i}^{2} + \frac {1}{2}
\sum_{\alpha,\beta =3,4} \sum_{i\neq j}^{N_\alpha,N_\beta}
V(r_{ij}) \, , \label{eq:H}
\end{equation}

 where the interaction is the same for all the different
 pairs of the mixture.

 A realistic, correlated, variational ground state wave function $\Psi_0$
 is obtained by the Jastrow--Feenberg ansatz \cite{Feenberg}

\begin{equation}
\Psi_0= F_J \, F_T \, F_{BF} \phi_0(N_3)\ ,
\label{eq:Psi_0}
\end{equation}
where $\phi_0(N_3)$ is the ground state Fermi gas wave function
for the $^3$He component and $F_J, F_T$ and $F_{BF}$ are N--body correlation
operators including explicit two--, three--body and back--flow dynamical
correlations respectively. We will limit our analysis to the case of
two--body, state independent (or Jastrow) correlations only
($F_T=F_{BF}=1$). $F_J$ results to be

\begin{equation}
F_J(N_4,N_3)=\prod_{i_3<j_3}^{N_3} f^{(3,3)}(r_{i_3 j_3}) \,
  \prod_{i_4<j_4}^{N_4} f^{(4,4)}(r_{i_4 j_4}) \,
  \prod_{i_3}^{N_3} \prod_{i_4}^{N_4} f^{(3,4)}(r_{i_3 i_4})
 \, . \label{eq:F_J}
\end{equation}
 where $f^{(\alpha,\beta)}(r)$ are two--body correlation functions
 determined by minimizing the variational ground state energy.

 It is possible to generate a correlated basis through
 the operator (\ref{eq:F_J}), to be used in a CBF perturbation theory
 (CBFPT). This theory has been succesfully adopted for computing the
 inclusive response of nuclear matter and heavy nuclei to electron
 and hadron scattering \cite{FabrociniRL,R_spin,RL_nuclei}, and has
 shown to be able to provide a semiquantitative agreement with
 experimental neutron inelastic scattering (nIS) data in pure, liquid
 atomic $^4$He \cite{Manousakis86}.

 In this paper, we will apply CBFPT to compute the nIS response of
 the mixture, by considering as intermediate states the normalized,
 correlated $^4$He n--phonon states (nPH) $|{\bf k}_1,..,{\bf k}_n\rangle$,
 and $^3$He n--particle, m--hole states (np--mh)
 $|{\bf p}_1,..,{\bf p}_n,{\bf h}_1,..,{\bf h}_m,\rangle$.

 The nPH states are given by

\begin{equation}
|{\bf k}_1,...,{\bf k}_n\rangle =
\frac {\rho_4({\bf k}_1)..\rho_4({\bf k}_n) |\Psi_0\rangle }
 {\langle \Psi_0 |\rho_4^\dagger({\bf k}_n)..\rho_4^\dagger({\bf k}_1)
 \rho_4({\bf k}_1)..\rho_4({\bf k}_n) |\Psi_0\rangle^{1/2} } \, ,
\label{eq:nPH}
\end{equation}
  where $\rho_4({\bf k})$ is the $^4$He density fluctuaction operator

 \begin{equation}
 \rho_4({\bf k})=\sum_{i=1,N_4} e^{ i{\bf {k \cdot r_i}}} \, .
\label{eq:rho4}
\end{equation}

Correlated np--mh states are obtained in a similar way, by applying the
correlation operator to the Fermi gas excited states $\Phi_{np-mh}(N_3)$,

\begin{equation}
 |{\bf p}_1,..,{\bf p}_n,{\bf h}_1,..,{\bf h}_m\rangle=
\frac {F_J|\Phi_{np-mh}\rangle }
{\langle \Phi_{np-mh}|F^\dagger_JF_J|\Phi_{np-mh}\rangle ^{1/2} } \, .
\label{eq:npmh}
\end{equation}

 We will consider 1--phonon (1PH) and 1p--1h intermediate correlated
 states, which we will term as One Intermediate Excitation (OIE) states.
 The response computed at the OIE
level will be called {\sl variational}. In addition, we will also consider
the possible decay of 1PH states into 2PH ones, which is essential in
giving a physically meaningful $^4$He excitation spectrum and provides
a quenching of the one--phonon peak.
This term will be computed in a boson--boson approximation,
{\sl i.e.} neglecting the $^3$He antisymmetry. Such an
approach may be justified on the basis of the low $^3$He concentration.

 1p--1h states may also be coupled to 1PH and 2PH. Such a coupling
 may be taken into account by a corresponding self--energy insertion.
Its analogous in the problem of the single $^3$He atom in $^4$He is
responsible for the impurity large effective mass. To estimate the importance
of this effect we will use the on--shell part of the impurity self--energy,
again relying on the small value of $x_3$.

 The plan of the paper is as follows. In section II we will briefly
outline the CBFPT for the response of the mixture and the variational
calculation will be described in some details. Section III is devoted
to the description of the calculation of the coupling with the 2PH states
and of the decays into 1PH and 2PH states. Section IV contains results for the
response and the comparison with the experimental scattering functions.
Moreover, the $^4$He and $^3$He excitation spectra are
presented and discussed. Conclusions are drawn in section V.

{\bf 2. CBFPT FOR THE RESPONSE}

The Dynamical Structure Function (DSF) $S(q,\omega)$ of a
$^3$He--$^4$He mixture at $T=0$ is given by the imaginary part of the
polarization propagator $D(q,\omega)$

\begin{equation}
S(q,\omega)=\frac {1}{\pi}
\Im D(q,\omega) ,
\label{eq:DSF}
\end{equation}
where
\begin{equation}
D(q,\omega)=\frac {1}{N}
\langle \tilde{\Psi}_0 | \rho^{\dagger}({\bf q})
\frac {1}{H-E_0-\omega -i \eta}
\rho({\bf q}) |\tilde{\Psi}_0 \rangle ,
\label{eq:POL}
\end{equation}
 and
\begin{equation}
\rho({\bf q})= \rho_3({\bf q})+\rho_4({\bf q}) ,
\label{eq:rho}
\end{equation}

\begin{equation}
\rho_{\alpha}({\bf q})=
\sum_{i=1,N_{\alpha}} e^{ i{\bf {q \cdot r_i}}}
\label{eq:rhoalpha}
\end{equation}

 and $N=N_3+N_4$. In eq.(\ref{eq:POL}), $\tilde{\Psi}_0$ is the exact
ground state of $H$ with eigenvalue $E_0$.

 The total DSF may be expressed in terms of partial $\alpha\beta$
DSF, $S_{\alpha \beta}(q,\omega)$,  as

\begin{equation}
S(q,\omega)=\sum_{\alpha,\beta=3,4}
x_{\alpha\beta}S_{\alpha \beta}(q,\omega) ,
\label{eq:DSF1}
\end{equation}
with $x_{\alpha\beta}=(x_\alpha x_\beta)^{1/2}$ and

\begin{equation}
S_{\alpha \beta}(q,\omega)=\frac {1}{\pi}
\Im D_{\alpha \beta}(q,\omega)= \frac {1}{\pi} \Im
\frac {1}{\sqrt{N_\alpha N_\beta }}
\langle \tilde{\Psi}_0 | \rho_{\alpha}^\dagger({\bf q})
\frac {1}{H-E_0-\omega -i \eta}
\rho_{\beta}({\bf q}) |\tilde{\Psi}_0 \rangle .
\label{eq:DSF2}
\end{equation}

 The experimentally measured nIS double differential cross section for the
 mixture directly provides access to the the total scattering function
 $\hat S(q,\omega)$, wich is in turn related to the partial DSFs' by
 the relation

\begin{equation}
\hat S(q,\omega)=\frac {x_4 \sigma_4 S_{44}(q,\omega) +
x_{34} \sigma_{34} S_{34}(q,\omega) +
x_3 \left[ \sigma_3 S_{33}(q,\omega)+ \sigma_3^i S_{33}^i(q,\omega)
\right] }
{x_4 \sigma_4 + x_3 ( \sigma_3^c + \sigma_3^i )} .
\label{eq:TSF}
\end{equation}

The elementary cross sections, as given by Sears \cite{Sears86} in units
of barns, are $\sigma_4=1.34, \sigma_3=4.42, \sigma_3^i=1.19$ and
$\sigma_{34}=4.70$. The inchoerent DSF $S_{33}^i(q,\omega)$ also appears
in the expression above, with the corresponding cross section $\sigma_3^i$.
Because of the absence of spin correlations, it describes $^3$He
spin fluctuactions via the operator

\begin{equation}
\rho_3^{\bf I}({\bf q})=
\sum_{i=1,N_3} e^{ i{\bf {q \cdot r_i}}} {\bf I}_i ,
\label{eq:rho3I}
\end{equation}

 where ${\bf I}_i$ is the spin of $^3$He i--nucleus.

 We will focus, in the remainder, mainly on the calculation of
$S_{\alpha \beta}$. To derive a
perturbative expansion it is convenient to split $H$ into an unperturbed
piece $H_0$ and an interaction term $H_1$, as follows

\begin{equation}
\langle m | H_0 | n \rangle = \delta_{nm} \langle m | H | m \rangle =
 E^v_{m} ,
\label{eq:H0}
\end{equation}
 and

\begin{equation}
\langle m | H_1 | n \rangle = ( 1 - \delta_{nm}) \langle m | H | n \rangle =
 \tilde{H}_{mn} .
\label{eq:H1}
\end{equation}

 Here $|m\rangle$ are correlated basis states, eigenstates of $H_0$.
In particular, $| 0 \rangle = | \Psi_0 \rangle $ is not an eigenstate of
$H$ and its difference from $| \tilde{\Psi}_0 \rangle$
is treated perturbatively.
The expansion is obtained by writing

\begin{equation}
H-E_0=H_0 - E^v_0 + ( H_1 - \Delta E_0 ) ,
\label{eq:HH0}
\end{equation}
where $\Delta E_0$ is the correction to the variational
ground state energy $E^v_0$, and by developing the propagator
$(H-E_0-\omega - i \eta)^{-1}$ in powers of $( H_1 - \Delta E_0 )$. A
similar expansion is performed for the ground state $| \tilde{\Psi}_0\rangle $.

 If the expansion is truncated at the zeroth order, the partial DSF
are given by:

\begin{equation}
S_{\alpha \beta}(q,\omega)
= \frac {1}{\sqrt{N_\alpha N_\beta }} \sum_n
\langle \Psi_0 | \rho_{\alpha}^\dagger({\bf q}) | n \rangle
\langle n | \rho_{\beta}({\bf q}) | \Psi_0 \rangle
\delta(\omega - \omega _n) ,
\label{eq:DSF3}
\end{equation}
with $\omega_n=E^v_n - E^v_0$.

 As stated in the introduction, we will first consider only OIE insertions,
{\sl i.e.} correlated 1PH and 1p--1h intermediate states, defined as:

\begin{equation}
|{\bf k}\rangle =
\frac {\rho_4({\bf k}) |\Psi_0\rangle }
 {\langle \Psi_0 |\rho_4^\dagger({\bf k})
 \rho_4({\bf k}) |\Psi_0\rangle^{1/2} } \, ,
\label{eq:1PH}
\end{equation}

\begin{equation}
 |{\bf p},{\bf h}\rangle=
\frac {F_J|\Phi_{1p-1h}\rangle }
{\langle \Phi_{1p-1h}|F^\dagger_JF_J|\Phi_{1p-1h}\rangle ^{1/2} } \, .
\label{eq:1p1h}
\end{equation}

{\sl 2.1 THE VARIATIONAL RESPONSES}

 The variational response is given by the sum of two components,

\begin{equation}
S_{\alpha \beta}(q,\omega)=S_{\alpha \beta}^{1PH}(q,\omega)+
S_{\alpha \beta}^{1p-1h}(q,\omega) ,
\label{eq:Ssum}
\end{equation}
 where $S_{\alpha \beta}^{1PH}(q,\omega)$ has a 1PH intermediate state

\begin{equation}
S_{\alpha \beta}^{1PH}(q,\omega)
= \frac {1}{\sqrt{N_\alpha N_\beta }} \sum_k
\langle \Psi_0 | \rho_{\alpha}^\dagger({\bf q}) | {\bf k} \rangle
\langle {\bf k} | \rho_{\beta}({\bf q}) | \Psi_0 \rangle
\delta(\omega - \omega _k) ,
\label{eq:S_PH}
\end{equation}
 and $S_{\alpha \beta}^{1p-1h}(q,\omega)$ has a 1p--1h intermediate state

\begin{equation}
S_{\alpha \beta}^{1p-1h}(q,\omega)
= \frac {1}{\sqrt{N_\alpha N_\beta }} \sum_{p,h}
\langle \Psi_0 | \rho_{\alpha}^\dagger({\bf q}) | {\bf p},{\bf h} \rangle
\langle {\bf p},{\bf h} | \rho_{\beta}({\bf q}) | \Psi_0 \rangle
\delta(\omega - \varepsilon_p+\varepsilon_h) .
\label{eq:S_1p1h}
\end{equation}
 $\omega_k$ and $\varepsilon_p-\varepsilon_h$ are the variational energies
of the OIE states considered.

$\omega_k$ is given by

\begin{equation}
\omega_k=
 \frac {1}{N_4}
\frac {\langle {\bf k}| H - E_0^v | {\bf k} \rangle}
{\langle {\bf k}| {\bf k} \rangle} =
 \frac {\hbar^2k^2}{2m_4S_{44}(k)} ,
\label{eq:wF}
\end{equation}
 and corresponds to the well known Feynman spectrum \cite{Feynman}. In
eq.(\ref{eq:wF}), $S_{44}(k)$ is the variational estimate of the 44
component of the Static Structure Function (SSF), $S_{\alpha\beta}(k)$,
given by:

\begin{equation}
S_{\alpha\beta}(k)=
 \frac {1}{\sqrt{N_\alpha N_\beta}}
\frac {\langle \Psi_0|\rho_{\alpha}^\dagger({\bf k}) \rho_{\beta}({\bf k})
 | \Psi_0 \rangle}
{\langle \Psi_0 | \Psi_0 \rangle} .
\label{eq:SSF}
\end{equation}

In a similar way, $\epsilon_{x=p,h}$ is obtained by

\begin{equation}
\epsilon_x=
\frac {\langle {\bf x}| H - E_0^v | {\bf x} \rangle}
{\langle {\bf x}| {\bf x} \rangle} ,
\label{eq:wph}
\end{equation}

 where $|{\bf x}\rangle$ is a particle or hole correlated state. We will
discuss later the evaluation of $\epsilon_x$.

By using the definition of the SSF given in eq.(\ref{eq:SSF}),
$\xi_\alpha({\bf q};{\bf k})=
\langle \Psi_0 | \rho_{\alpha}^\dagger({\bf q}) | {\bf k} \rangle$ is readily
obtained as

\begin{equation}
\xi_\alpha({\bf q};{\bf k})
= {\sqrt{N_{\alpha}}}\frac {S_{\alpha 4}(k) }{\sqrt{S_{44}(k)}}
\delta_{{\bf k}-{\bf q}} ,
\label{eq:csi}
\end{equation}

 giving, for $S_{\alpha \beta}^{1PH}(q,\omega)$,

\begin{equation}
S_{\alpha \beta}^{1PH}(q,\omega)
= \sum_k \frac
{S_{\alpha 4}(k) S_{4 \beta }(k)}{S_{44}(k)}
\delta_{{\bf k}-{\bf q}} \delta(\omega - \omega _k) .
\label{eq:S_PH1}
\end{equation}

 The one--phonon contribution to the variational $\alpha$--$\beta$
responses shows a delta--like behavior, whose strenght is
$Z_{\alpha\beta}^v(k)=S_{\alpha 4}(k) S_{\beta 4}(k)/S_{44}(k)$,
and it is located at the Feynman phonon energy. We notice that ({\sl i})
$Z_{44}^v(k)=S_{44}(k)$ and that ({\sl ii}) the 33 and 44 variational
DSF are positive ($S_{44}(k)$ being positive), whereas this may not
be true for the 34 DSF.

  The expression for the particle--hole response
  $S_{\alpha \beta}^{1p-1h}(q,\omega)$ is more involved. A detailed
  description for a pure Fermi system (specifically, nuclear
  matter) can be found in \cite{FabrociniRL} and references therein.
  On the basis of that formalism, the extension to a boson--fermion
  mixture is straightforward.

  In CBF theory, the non diagonal matrix elements
  $\xi_\alpha({\bf q};{\bf p},{\bf h})=
\langle \Psi_0 | \rho_{\alpha}^\dagger({\bf q}) | {\bf p},{\bf h} \rangle$ are
computed by a cluster expansion in Mayer like diagrams, and by
summing infinite classes of relevant terms. The
  $\xi_\alpha$ are explicitely given by:

\begin{equation}
\xi_\alpha({\bf q};{\bf p},{\bf h})
=
\delta_{{\bf q}-{\bf p}+{\bf h}} \frac {1}{\sqrt{D(p)D(h)}}
( \tilde{h}_{dd,\alpha 3}(q) + \delta_{\alpha 3} [ 1 +
 \tilde{h}_{ed,33}(q) ] ) ,
\label{eq:csiph}
\end{equation}
 where

\begin{equation}
\tilde{h}_{xy,\alpha 3}(q)
= \rho_\alpha \int d^3r e^{\imath {\bf q}\cdot {\bf r}}
[ g_{xy,\alpha 3}(r) - \delta_{xy,dd} ] ,
\label{eq:hxy}
\end{equation}
 with $(x,y)=(d,e)$ and $g_{xy,\alpha \beta}(r)$ are partial radial
 distribution functions (RDF). In fact, the total $\alpha \beta$--RDF,
 $g_{\alpha \beta}(r)$, giving the probability of finding a
 $\alpha$--type particle 1 at a distance $r_{12}$ from a
 $\beta$--type particle 2,

\begin{equation}
 g_{\alpha \beta}(r_{12})
= \frac {N_\alpha ( N_\beta - \delta_{\alpha \beta} )}
        {\rho_\alpha \rho_\beta}
  \frac {\displaystyle \int d^3r_3 .. d^3r_N \vert \Psi_0 \vert ^2}
        {\displaystyle \int d^3r_1 .. d^3r_N \vert \Psi_0 \vert ^2}  ,
\label{eq:RDF}
\end{equation}
 is computed, in Fermi Hypernetted Chain (FHNC) \cite{Boronat93}, using the
 correlated g.s. $\Psi_0$ and it turns out to be written as:

\begin{equation}
 g_{\alpha \beta}(r_{12})
 = g_{dd,\alpha \beta}(r_{12}) + \delta_{\beta 3} g_{de,\alpha 3}(r_{12}) +
   \delta_{\alpha 3} g_{ed,3 \beta}(r_{12}) +
   \delta_{\alpha 3} \delta_{\beta 3} g_{ee,33}(r_{12}) .
\label{eq:RDF1}
\end{equation}

  The partial RDF are classified, in FHNC theory, according to whether
  the external particle (1 or 2) is reached by a statistical correlation
  ({\sl i.e.} if the particle is involved in an exchange loop, $e$--vertex),
  or by a dynamical correlation, $(f^{(\alpha,\beta)})^2 - 1$, only
  ($d$--vertex).

  The definitions of the partial RDF, together with the full set of
  the related FHNC equations, may be found in \cite{Boronat93}.

  Actually, eq.(\ref{eq:csi}) sums all cluster diagrams factorizable in
  products of dressed, two--body diagrams. They do not contain only
  two--body cluster terms, but include, in turn, an infinite
  number of particles, as they are written in terms of the RDFs, rather than
  the bare two--body correlations.

  Three--body, non factorizable diagrams are also present in the cluster
  expansion of $\xi_\alpha$, even if they do not appear in eq.(\ref{eq:csi}).
  However, they have been inserted, following ref.\cite{FabrociniRL}.

  The function $D(x=p,h)$ is

\begin{equation}
 D(x) = 1 -  \rho_3 \int d^3r e^{\imath {\bf x}\cdot {\bf r}}
( g_{dd,3 3}(r) - 1 ) L(k_F r) ,
\label{eq:D_x}
\end{equation}
 where $L(k_F r)$ is the FHNC  generalization of the exchange Slater
 function  $l(k_F r)= 3 j_1(k_F r)/(k_F r)^3$, and $k_F$ is the $^3$He
 Fermi momentum ($k_F^3=3 \pi^2 \rho_3$).

 Again, as $D(x)$ turns out to be positive, $S_{33,44}^{1p-1h}(q,\omega)$
 are positive, whereas $S_{34}^{1p-1h}(q,\omega)$ may be not.

 Finally, the spin fluctuaction matrix element,
  $\xi_3^{\bf I}({\bf q};{\bf p},{\bf h})= \langle \Psi_0 |
  \rho_3^{{\bf I}\dagger}({\bf q}) | {\bf p},{\bf h} \rangle$, is simply
  given by:

\begin{equation}
\xi_3^{{\bf I}}({\bf q};{\bf p},{\bf h})
=
\delta_{{\bf q}-{\bf p}+{\bf h}} \frac {1}{\sqrt{D(p)D(h)}} .
\label{eq:csiphI}
\end{equation}

{\bf 3. CORRELATED ONE AND TWO PHONON INTERMEDIATE STATES }

In this section we will first study the effect on the phonon responses
of the insertion of orthogonal, correlated 2PH states:

\beq
|{\bf k}_1 {\bf k}_2\rangle_o = \left( 1-|{\bf k}\rangle\langle{\bf k} |\right)
|{\bf k}_1 {\bf k}_2\rangle ,
\eeq
 where the 2PH states of eq.(\ref{eq:nPH}) have been orthogonalized
 to the 1PH ones by a Gram-Schmidt procedure.

2PH states influence the partial polarization propagators
$D_{\alpha \beta}(q,\omega)$ via the direct coupling to the ground
state and via the decay of 1PH states into 2PH. The coupling to the g.s
goes through the matrix element of the $^3$He fluctuaction operator,

\beq \label{csi3}
\xi_3({\bf q};{\bf k}_1,{\bf k}_2)=
\langle \Psi_0|\rho_3^{\dagger}({\bf q})|{\bf k}_1,{\bf k}_2\rangle_o ,
\eeq

( notice that $\xi_4({\bf q};{\bf k}_1,{\bf k}_2)$ vanishes because of the
Schmidt
orthogonalization of the 2PH states), whereas the decay is driven by
the non diagonal matrix element of the hamiltonian

\beq \label{a}
a({\bf k};{\bf k}_1,{\bf k}_2)=\langle {\bf k}|H_1|{\bf k}_1,{\bf k}_2
\rangle_o .
\eeq

 These CBF matrix elements have been computed
 in a boson--boson approximation (treating the $^3$He as a mass--3
 boson) and by adopting the Convolution Approximation (CA) for the
 three--body distribution functions.

 Their explicit expressions are:

\begin{equation}
\xi_3({\bf q};{\bf k}_1,{\bf k}_2)=
\frac{\displaystyle S_{34}(k_1)S_{34}(k_2)}
{\displaystyle \sqrt{S_{44}(k_1)S_{44}(k_2)}}\left(S_{33}(q)-\frac{S_{34}^2(q)}
{S_{44}(q)}
\right),
\end{equation}

 and

\begin{eqnarray}
a({\bf k};{\bf k}_1,{\bf k}_2)& = &
\frac{\hbar ^2}{\sqrt{N_4}2m_4}\left(
\frac{{\bf k}\cdot{\bf k}_1S_{44}(k_2)+{\bf k}\cdot{\bf k}_2S_{44}(k_1)-
k^2S_{44}(k_1)S_{44}(k_2)}
{\sqrt{S_{44}(k)S_{44}(k_1)S_{44}(k_2)}}-\right. \\ \nonumber & &
\left. \sqrt{\frac{x_4}{x_3}}\frac{k^2S_{34}(k)S_{34}(k_1)S_{34}(k_2)}
{S_{44}(k)\sqrt{S_{44}(k)S_{44}(k_1)S_{44}(k_2)}}
\right) .
\end{eqnarray}

It is convenient, at this point, to introduce the correlated
 self-energy

\beq
\Sigma_1(k,\omega)=\frac{1}{2}\sum_{{\bf k}_1,{\bf k}_2}
\frac {|a({\bf k};{\bf k}_1,{\bf k}_2)|^2}{\omega_{k_1}+\omega_{k_2}-\omega-
i\eta} ,
\eeq

 and the function $\chi({\bf q};k,\omega)$ given by:

\beq
\chi({\bf q};k,\omega)=\frac{1}{2}\frac{1}{\sqrt{N_3}}\sum_{{\bf k}_1,{\bf
k}_2}
a({\bf k};{\bf k}_1,{\bf k}_2)
\frac{1}{\omega_{k_1}+\omega_{k_2}-\omega- i\eta}
\xi^{\dagger}_3({\bf q};{\bf k}_1,{\bf k}_2) .
\eeq

 If we define the dressed phonon propagator $G^d(k,\omega)$ as

\beq
G^d(k,\omega)=\frac{1}{\omega_k-\Sigma_1(k,\omega)-\omega -i\eta} ,
\eeq

 then, the phonon contributions to the polarization propagators can
 be rearranged as:

\beq
D^{PH}_{44}(q,\omega)=\frac {1}{N_4}
\sum_{\bf k} |\langle \Psi_0|\rho_4^{\dagger}({\bf q})|{\bf k}\rangle|^2
G^d(k,\omega),
\eeq

\beq
D^{PH}_{34}(q,\omega)=
\frac {1}{\sqrt{N_3N_4}}
\sum_{\bf k} \langle \Psi_0|\rho_4^{\dagger}({\bf q})|{\bf k}\rangle
G^d(k,\omega)
[ \langle {\bf k} |\rho_3({\bf q})|\Psi_0\rangle + \frac {1}{\sqrt{N_4}}
\chi({\bf q};k,\omega) ] ,
\eeq

 and

\beq
D^{PH}_{33}(q,\omega)=
\frac {1}{N_3}
\sum_{\bf k} \langle \Psi_0|\rho_3^{\dagger}({\bf q})|{\bf k}\rangle
G^d(k,\omega)
[ \langle {\bf k} |\rho_3({\bf q})|\Psi_0\rangle + \frac {2}{\sqrt{N_3}}
\chi({\bf q};k,\omega) ] .
\eeq

 The DSF are then obtained by taking the imaginary
 parts of $D_{\alpha \beta}$ . Terms quadratic in
 $\xi_3({\bf q};{\bf k}_1,{\bf k}_2)$
 have not been considered.

The relevant changes introduced by the insertion of the 2PH states
in the phonon responses are:

\begin{enumerate}

\item the strengths of the delta--like 1PH peaks $Z_{\alpha \beta}$
are  generally quenched respect to $Z^v_{\alpha \beta}$. In the
44 case, we have

\beq
Z_{44}(k)=Z^v_{44}(k) \left(
1 + \frac {\partial \Re \Sigma_1(k,\omega )}{\partial \omega}
\right)_{\omega = \omega_k}^{-1} .
\eeq

 Analogous corrections occur for $Z_{34}(k)$ and  $Z_{33}(k)$, which are
 also affected by those parts
 of the polarization propagators containing $\xi_3({\bf q};{\bf k}_1,{\bf
k}_2)$;

\item the 1PH peaks are shifted by the real part of the on--shell
 self energy, since the $^4$He spectrum is modified as

\beq
\omega_k \longrightarrow \omega^{CBF}_k = \omega_k +
\Re \Sigma_1(k,\omega^{CBF}_k) ;
\eeq

\item a multiphonon tail appears at large $\omega$--values, beyond
 the position of the 1PH peak, at the momentum transfers here
 considered.

\end{enumerate}

{\bf 4. CBF RESPONSES}

In the class of the Jastrow correlated wave functions, the best
variational choice is provided by the solution of the Euler equations

\beq
 0 =
\frac {\delta \langle \Psi_0| H | \Psi_0 \rangle }
{\delta f^{(\alpha \beta)}} .
\eeq

 The resulting equations have been derived, within the FHNC framework,
 and solved for the $^3$He impurity problem \cite{Owen81,FP84}, for
 the boson--boson mixture \cite{Chakraborty83} and, lately, for the
 real fermion--boson case \cite{Saarela_rep}.

 Another, often used approach consists in parametrizing the correlation
 functions and in minimizing the ground state energy with respect to
 the parameters.
 This is the choice we have adopted here. Besides that, some of the
 results we will present have been obtained within the Average
 Correlation Approximation (ACA). In ACA, the correlation functions
 are the same for all the types of pairs ($f^{(3,3)}=f^{(3,4)}=f^{(4,4)}$)
 and the differences in the distribution functions (or in
 the static structure functions) are due only to the different isotope
 densities and statistics. We will also show that going beyond the
 ACA does not affect our results.

 We have used three types of correlation functions: the time honored, short
 ranged McMillan form (SR) and two long ranged functionsa
 (LR and LR1).

 The McMillan correlation, in ACA, is given by:

\beq
 f_{SR}(r) =
 exp \left[- \left( \frac {b \sigma}{r}\right)^5
      \frac {1}{2} \right] ,
\eeq

 where $b=1.18$ and $\sigma=2.556 \AA$. The SR correlation function
 gives a good description of the short range behavior of the pair
 wave function but fails to reproduce long range properties. For
 istance, it does not ensure the linear behavior of the $^4$He
 SSF at $k \rightarrow 0$ (the phonon dispersion).
 Such a dispersion reflects in a long range behavior
 of the correlation of the type $f(r\rightarrow \infty) - 1 \propto
 -r^{-2}$. To this aim, we have also used a modified form,
  having the correct long range structure (LR), given by

\beq
 f_{LR}(r) = f_{SR}(r)
 \left[ A + B exp
  \left( - \frac {(r-D)^2}{\tau r^4} \right) \right].
\eeq

 The parameters of $f_{LR}(r)$, giving the variational mimimum of the
 $^4$He energy at the equilibrium density $\rho_0=0.02185 \AA ^{-3}$,
 are $b=1.18, A=0.85, B=1 - A, D=3.8 \AA$ and $\tau= 0.043 \AA ^{-2}$
 (see Ref.\cite{Boronat93} for more details about the energetics
 of the mixture). The $B$ and $\tau$ parameters are related to the
 experimental pure $^4$He sound velocity $c$ and to the low--$k$ behavior
 of its SSF by the relations:

\beq
 \frac {B}{\tau} =
  \frac {m_4 c}{2 \pi ^2 \hbar \rho_0} \,;\,
  S^{(4,4)}(k \rightarrow 0) = \frac {\hbar k}{2 m_4 c} .
\eeq

In order to check the accuracy of ACA, we have also used
a LR correlation (LR1), formally identical to $f_{LR}$, but with parameters
 depending on the type of the correlated pair.
The 44 correlation function is the same as above, whereas the
parameters of the 43 and 33 ones have been obtained by minimizing
the energy of the pure $^4$He with one and two $^3$He impurities,
respectively \cite{Boronat93}.

 Key ingredients in the CBF theory of the response in Helium mixtures
are the radial distribution functions $g_{\alpha \beta}(r)$ and the
static structure functions $S_{\alpha \beta}(k)$. Figs.(1) and (2)
show these quantities in a $4.7\%$ mixture, at a total density
$\rho=0.02160 \AA ^{-3}$,
for the $f_{LR1}(r)$ correlation, in FHNC/0
approximation ($i.e.$ we have neglected the elementary diagrams
\cite{Boronat93}). The results for the SSF, with the $f_{SR}(r)$,
differ mainly in the region of low--$k$ values, in agreement with the
previous discussion.

 Table (1) shows the variational strenghts $Z^v_{\alpha \beta}(k)$ of the
 one--phonon response for the same mixture and compares the results obtained
 with the SR and LR  correlation functions at four momentum values, from
$q=1.1$
 to  $1.7 \AA^{-1}$. The positions of the variational delta peaks,
 $\omega_k$, are also given. It has to be noticed that the Feynman
  spectrum overestimates the experimental data by at least $10 K$ both
 in the maxon and roton regions.

 Table (2) provides the same quantities after the insertion of the
 2PH states. Fig.(3) shows the $^4$He spectrum with the LR1 correlation.
 The figure also compares the spectrum with pure $^4$He at $\rho_0$
 and with the experimental results of ref.\cite{Fak90} (circles) in a
 $x_3=1.1 \%$ mixture at SVP and of ref.\cite{Hilton77} (squares)
 for a $x_3=6.0 \%$ mixture.

 The changes in going from pure $^4$He to the mixture are clearly
 visible.
 This is mainly a density effect. In fact, we obtain
 similar results if the LR--ACA correlation is used.
 CBF perturbative corrections appear to be large and bring
 the maxon energy close to the experiments. The roton is not well
 described, as it is too shallow respect to the data. This feature
 is also present in the $^4$He case. As for the $^3$He spectrum, we
 believe that most of the discrepancy in this part of the spectrum has
 to be ascribed to the use of CA in the
 calculation of the CBF matrix elements. Moreover, contributions
 from higher order CBF pertubative diagrams are known to be important to
 correctly reproduce the roton minimum in pure $^4$He \cite{Lee75}.
 However, the CA results show
 a change in the sign of the shift from mixture to pure system
 at $q\simeq 1.8 \AA^{-1}$, in good agreement with the measured
 experimental value \cite{Fak90} at constant pressure
($q\simeq 1.9 \AA^{-1}$).
 The boundaries of the
 $1p-1h$ DSFs are related to the energies of the $1p-1h$ state
 $\epsilon_p-\epsilon_h$. $\epsilon_{p(h)}$ has been computed
 by the procedure of ref.\cite{VRP_eps}. However, because
 of the low $^3$He density, it turns out to be extremely close to
 the free Fermi gas spectrum

\begin{equation}
\epsilon_k^{FG}=
 \frac {\hbar^2k^2}{2m_3} ,
\label{eq:epsFG}
\end{equation}

 so, $\epsilon_k^{FG}$ has been used in all the calculations.
 Perturbative corrections to $\epsilon_k$ may be computed in CBF.
 In the case of the $^3$He impurity, CBFPT provides an accurate evaluation
 of its spectrum \cite{Fabrocini86,Fabrocini_prep} if the decay of the
 impurity excited state (given by a correlated plane wave)
 into correlated 1PH and 2PH states is considered. 1PH states account
 for $\sim 2/3$ of the difference between the experimental effective mass and
 the bare one, whereas 2PH states give the remainder. Because of the low
 density of the $^3$He component, it is reasonable to expect a similar
 behavior in the finite concentration mixture. It implies that we should
 insert in the CBFPT expansion the coupling between $1p-1h$ states and 1PH
 and 2PH ones.

 Work along this line is in progress. Here we have used for
 $\epsilon_k^{CBF}$ the CBFPT spectrum of the single impurity,
 obtained by extending to finite momenta
 the approach of  ref.\cite{Fabrocini86} for the effective mass.
 The involved matrix elements have been computed in CA
  for the three--body distribution functions.
 CA gives $m_3^*(CA)\sim 1.8 m_3$ for the impurity, whereas the more realistic
 Superposition Approximation (SA) gives $m_3^*(SA)\sim 2.2 m_3$. However,
 the SA $k\neq 0$ matrix elements are much more involved than their CA
 counterparts, and their evaluation, together with a description of the method,
  will be presented subsequently \cite{Fabrocini_prep}.  Here, the
 effect of the missing effective mass has been estimated by simply
 scaling the CA spectrum as $\epsilon_k^{CBF}=[m_3^*(CA)/m_3^*(expt)]
 \epsilon_k^{CBF}(CA)$.

 Fig.(4) shows the $^3$He spectra in different approximations, and compares
 them with the experimental data (circles from ref.\cite{Fak90} and
 squares from ref.\cite{Hilton77}) and with the LP and LPM parametrizations
 given in the introduction, with parameters $m_3^*=2.3 m_3$ and
 $\gamma = 0.132 \AA^2$. Even from the CA calculation,
 a deviation from the LP behavior clearly appears  .
 The estimated CBF value of $\gamma$ in CA turns out to be
 $\gamma(CA)=0.052 \AA^2$. We stress once more that we expect SA to
 provide a better description of the $^3$He spectrum behavior,
 as it correctly takes into account the core property of the system,
 requiring that the three--particle distribution functions vanish
 when any interparticle distance is lower than the radius of the
 repulsive core of the potential.

 Fig.(5) gives the $1p-1h$ DSF $S_{\alpha \beta}^{1p-1h}(q,\omega)$ at
 two momentum values ($q=1.3$ and $1.7 \AA^{-1}$), with the LR1 correlation
 and using the CBF  $^3$He spectrum. The two 33 DSF are
 very close and dominant, becoming indistinguishable at higher
 momenta; the 44
 component is always very small (notice that it has been amplified by
 a factor of 10 in the figure); the 34 part is negative and
 an order of magnitude larger than  $S_{44}^{1p-1h}$ in absolute value,
 contributing to decrease the total response mainly at low momenta.
 The free Fermi Gas DSF would be located to a larger energy  with a
 lower peak strength, compatible with the fact that the correlated
 system has a $^3$He effective mass 2.3  times larger than the bare mass
 (at $q=1.3 \AA^{-1}$ the FG peak position  is $\omega=13.6 K$ and the
 strength is $S_{33}^{1p-1h}(FG)=.115 K^{-1}$). In addition, as for the
 phonon DSF, the use of the SR and LR correlations does not alter
 appreciably the results shown in the figure.

 Table (3) shows the CBF values of the $m_{n,\alpha\beta}(q)$ sum rules
 of the DSF's, for $n=0,1$, defined as:

\begin{equation}
m_{n,\alpha\beta}(q)=
\int d\omega ~ \omega^n ~ S_{\alpha\beta}(q,\omega).
\label{eq:nSR}
\end{equation}

The {\sl exact} DSF's satisfy the $f-$sum rules

\begin{equation}
m_{1,\alpha\beta}(q)= \frac {\hbar^2 q^2}{2 m_\alpha} \delta_{\alpha\beta},
\label{eq:fSR}
\end{equation}

and for $n=0$ one has $m_{0,\alpha\beta}(q)=S_{\alpha\beta}(q)$.

The table gives also the variational values of the SSF's of eq.\ref{eq:SSF}
and the $f-$sum rules.
$m_0$ and the SSF's, as well as $m_1$ and the $f-$sum rules are in good
 agreement for all $q$'s in the 44 case. For 33, $m_0$ is reasonably close
to the variational SSF, whereas the $f-$sum rule is underestimated (we
recall that the SSF's have been computed with the {\sl variational} ground
states, not the exact one). The 34 $m_n$ are not satisfying, especially for
$m_1$.  In order to trace down
the source of the differences we also show the contributions to the sums from
the large--energy multiphonon tails of the DSF's, $m_{1,0}(mPH)$.
The tail provides a large part of the sum rules and, for $\alpha\beta=34$,
it is dominant. The effect is more clearly visible in $m_1$, where the tails
of the DSF's are multiplied by a large energy factor. As a consequence,
a quantitatively correct estimate of
$m_{1,\alpha\beta}$ would porbably require a more
accurate evaluation of the multiphonon contributions as well as
the insertion of higher order perturbative diagrams.

 To evaluate the total scattering function $\hat S(q,\omega)$, the
 DSF must be multiplied by the elementary cross sections and the
 concentrations of the species. In fig.s(6a,b) we give the partial CBF
 scattering functions (PSF):

\begin{equation}
 \hat S_{44}(q,\omega)=
 \frac {x_4\sigma_4}{x_4+x_3(\sigma_3^c+\sigma_3^i)}S_{44}(q,\omega),
\label{eq:PSF44}
\end{equation}
\begin{equation}
 \hat S_{34}(q,\omega)=
 \frac {x_{34}\sigma_{34}}{x_4+x_3(\sigma_3^c+\sigma_3^i)}S_{34}(q,\omega)
\label{eq:PSF3}
\end{equation}
and
\begin{equation}
 \hat S_{33}(q,\omega)=
 \frac {x_3}{x_4+x_3(\sigma_3^c+\sigma_3^i)}
 [\sigma_3S_{33}(q,\omega)+\sigma_3^iS_{33}^i(q,\omega)].
\label{eq:PSF33}
\end{equation}

 The LR1 correlation has been used. The position and the strength of
 the phonon contribution to the PSF's are explicitely given. In the
 $1p-1h$ sector, at the lower momentum, the 33 PSF is strongly
 reduced by the 34 PSF, which practically disappears at $q=1.7 \AA^{-1}$.
 The 44 PSF is always negligible in this sector. In the phonon
 sector, the 44 PSF is the dominant one. The 33 component always
 results to be very small. The 34 PSF at $q=1.3 \AA^{-1}$ slightly
 reduces the scattering function, while at $q=1.7 \AA^{-1}$ increases it.

 In order to compare with the experimental scattering function, the
 theoretical PSF's have to be convoluted with the experimental broadening
 functions. As at these momentum transfers the phonon peak is still
 delta shaped, because there is no overlap with the multiphonon background,
 we assume, in accordance with the authors of ref.\cite{Fak90}, that the
 width in energy of the low temperature results in that reference
 is entireky due to the instrumental resolution. For this reason we have
 convoluted the phonon peak  with a gaussian having an average
 half maximum width  of $1.3 K$. A gaussian with a width of $1.2 K$
 \cite{Fak_priv} has been used for the $1p-1h$ response.

 The convoluted total scattering functions are compared
 with the experimental results of ref.\cite{Fak90} in fig.s(7a,b,c)
 at $q=1.1$, $1.5$ and $1.7 \AA^{-1}$, respectively, for the $4.7\%$
 mixture we have considered so far.

 At $q=1.1 \AA^{-1}$, both the position and the strength of the
 phonon branch are well described by our calculation. When approaching
 the roton minimun region, the agreement worsens and we overestimate
 the experimental data. As discussed previously, we expect that the
 use of SA will improve the CBF description.

 An analogous analysis can be performed fort the $1p-1h$ sector.
 The use of the CBF--CA spectrum slightly misses the location of the
 bump, well described in turn by a LPM parametrization, which is
 essentially a fit to the experimental data. We recall that the relevant
 difference between the LPM and the CBF--CA energies lies in the
 $\gamma$--parameter value, smaller by a factor $\sim 0.4$ in the
 latter case. A simple, quadratic LP parametrization with $m_3^*=2.3 m_3$
 seems to be ruled out.

 The $^3$He scattering function $\hat S_3(q,\omega)$, defined as

\begin{equation}
 \hat S_{3}(q,\omega)=
 \frac {x_4\sigma_4+x_3(\sigma_3^c+\sigma_3^i)}{x_3(\sigma_3^c+\sigma_3^i)}
 \hat S(q,\omega),
\label{eq:SF3}
\end{equation}

and the function $\bar S_3(q,\omega)$, given by

\begin{equation}
 \bar S_{3}(q,\omega)=
 \frac
 {\sigma_3S_{33}(q,\omega)+\sigma_3^iS_{33}^i(q,\omega)}
{\sigma_3^c+\sigma_3^i} ,
\label{eq:SF3_3}
\end{equation}

 in the $1p-1h$ sector, are given in fig.(8). The figures contains also
 the convolution of $\hat S_3(q,\omega)$ with the experimental
 broadening function and the experimental results of ref.\cite{Fak90},
 at $q=1.3 \AA^{-1}$. $\hat S_3$ and $\bar S_3$  are identical if
 $S^{1p-1h}_{34}=S^{1p-1h}_{44}=0$. So, their differences are basically
 a measure of the importance of the 34 contribution (the 44 one being
 negligible).
 Our results show a large suppression of $S^{1p-1h}_{33}$ due to
 $S^{1p-1h}_{34}$, which brings the CBF response much closer to the
 experiments.

{\bf 5. CONCLUSIONS}

 Correlated Basis Perturbation Theory has been used to microscopically
 compute the scattering function in a $x_3=4.7\%$ $^3$He--$^4$He mixture at
 T=0.  The theory has allowed for explicitely separating the different
 contributions to the response and for semiquantitatively assessing the
 relevance of the 34 component.
 In the $1p-1h$ region, the $S_{33}$ response
 is sizeably reduced by $S_{34}$ up to $q\simeq 1.5 \AA^{-1}$,
 whereas $S_{44}$ is always negligible. A similar effect, even if smaller
 in magnitude, is present in the phonon--roton sector, where the dominant
 $S_{44}$ is only sligthly modified by $S_{34}$.

 The responses have been computed by inserting correlated $1p-1h$ and
 1-- and 2--phonon intermediate states. Also the possible decay of
 1--phonon into 2--phonon states has been estimated in boson--boson
 approximation and using the Convolution Approximation for the
 three--body distribution functions.

 The microscopic quasiparticle $^3$He spectrum clearly shows a deviation
 from the simple LP form. The spectrum has been actually computed for
 the single impurity problem, but we do not believe that its evaluation
 in the low concentration mixture will dramatically change our findings.
 In particular, a deviation from LP was advocated in ref.\cite{Fak90}
 to explain the experimental $1p-1h$ response, in contrast with a
 possible large change of the $^3$He effective mass in mixture
 (from $m_3^*=2.3 m_3$ at $x_3=0$ to $m_3^*=2.9 m_3$ at $x_3=4.7\%$).
 The CBF spectrum still does not reproduce fully quantitatively the data,
 and a more accurate calculation is needed.

 The $^4$He excitation spectra in the phonon--roton branch of the
 pure system and the mixture at SVP have been compared. The shift
 between the two excitations appears to be due to the change in
 density. CBF gives a good description  of the maxon region, but
 overestimates the roton, even if it gives an almost  correct
 $q$--value for the change of sign of the shift .

 The CBF scattering function at low momenta gives a reasonable
 description of the scattering data (both for the position and
 strength). The agreement worsens as $q$ increases.
 The peaks are located at too a large energy and
 their strength is overestimated.  We believe that the reason of
 this lies in the approximations made to compute the decay of 1PH
 states into 2PH and in the lack of higher intermediate states,
 which become more and more important as the momentum increases.
 In particular, the $1p-1h$ sector does not include two probably
 relevant contributions: the decays of $1p-1h$ states into
  (1) $2p-2h$ and (2) $1PH$ states. The former adds large
 energy tails to the $1p-1h$ bump reducing its strength, and the
 latter is known to be responsible for a large part
 of the $^3$He effective mass. Our CBF calculation includes
 the real part of the $1p-1h$ into $1PH$ decay but does not
 consider its imaginary part.

 The importance of the 34 contribution to the total scattering
 function is especially visible in the $^3$He scattering function
 in the $1p-1h$ region, where its introduction reduces the response
 by a factor $\sim 0.6$ at $q=1.3 \AA^{-1}$.

 More work is clearly needed in order to give a fully quantitative
 description of both the excitations and the responses of the Helium
 mixtures. However, from our results, we believe that CBF is a promising
 theory in view of achieving this goal.

{\bf AKNOWLEDGEMENTS}

The authors are grateful to Bjon F{\aa}k for several ftuiful exchanges.
This research was supported in part by DGICYT (Spain) Grant
Nos. PB92-0761, PB90-06131, PB90-0873 and the agreement DGICYT
(Spain)--INFN (Italy).

\begin{figure}
\caption{ Radial distribution functions for the mixture (see text).
The solid line gives $g_{44}$, the dashed line $g_{34}$ and the
dotted one is $g_{33}$.}
\label{fig:1}
\end{figure}

\begin{figure}
\caption{ Static structure functions for the mixture (see text).
The solid line gives $S_{44}$, the dashed line $S_{34}$ and the
dotted one is $S_{33}$.}
\label{fig:2}
\end{figure}

\begin{figure}
\caption{ $^4$He excitation spectrum in the mixture (crosses) and in the
pure system (dashed line). The upper curves are the Feynam spectra.
Squares and circles are mixture experimental data (See text).}
\label{fig:3}
\end{figure}

\begin{figure}
\caption{ $^3$He excitation spectrum. The dashed line gives
$\epsilon_k^{CBF}(CA)$, the solid line is $\epsilon_k^{CBF}$;
also shown are the LP, LPM and free (F) spectra.
Squares and circles are the experimental data (See text).}
\label{fig:4}
\end{figure}

\begin{figure}
\caption{ $1p-1h$ DSF at $q=1.3$ and $1.7 \AA^{-1}$. The
 continuous line gives  $S_{33}$,the dot--dashed $S_{33}^i$, the dotted
$10\times S_{44}$ and the dashed $S_{34}$.}
\label{fig:5}
\end{figure}

\begin{figure}
\caption{ CBF Partial scattering functions at $q=1.3$ (6a) and $1.7 \AA^{-1}$
 (6b). Continuous line $\hat S_{33}$, dashed line $\hat S_{34}^i$,
 dotted $\hat S_{44}$. The PH--$\alpha\beta$ numbers are the strengths
 of the phonon PSF, located at $\omega_q$.}
\label{fig:6}
\end{figure}

\begin{figure}
\caption{ Total scattering functions at $q=1.1$ (7a), $q=1.5$ (7b)
and $1.7 \AA^{-1}$ (7c) (solid lines). Also shown are the
$1p-1h$ responses with the LP (dotted lines) and LPM (dashed lines)
spectra and the experimental data (crosses).}
\label{fig:7}
\end{figure}

\begin{figure}
\caption{ $1p-1h$ $^3$He scattering function at $q=1.3 \AA^{-1}$.
The solid line is $\hat S_3(q,\omega)$, the dashed line is
$\bar S_3(q,\omega)$, the dotted line is the experimental convolution
of $\hat S_3(q,\omega)$  and crosses are the experimental data.}
\label{fig:8}
\end{figure}

\begin{table}[]
\caption{Variational strenghts and positions of the one phonon DSF responses
with different correlations (see text). $q$ in $\AA^{-1}$ and
$\omega_q$ in $K$}
\begin{tabular}{lccccc}
        &$q$ & $\omega_q$ & $Z^v_{44}$ & $Z^v_{34}$ & $Z^v_{33}$ \\ \hline
 $ SR $&  1.1 & 20.44      & 0.356      & -0.146     & 0.060       \\
 $ LR $&      & 20.77      & 0.361      & -0.145     & 0.058       \\
 $ LR1$&      & 20.75      & 0.362      & -0.147     & 0.060       \\
 $ SR $&  1.3 & 20.74      & 0.491      & -0.115     & 0.027       \\
 $ LR $&      & 21.68      & 0.461      & -0.122     & 0.032       \\
 $ LR1$&      & 21.71      & 0.460      & -0.122     & 0.032       \\
 $ SR $&  1.5 & 19.91      & 0.681      & -0.072     & 0.008       \\
 $ LR $&      & 19.99      & 0.686      & -0.071     & 0.007       \\
 $ LR1$&      & 20.02      & 0.685      & -0.069     & 0.007       \\
 $ SR $&  1.7 & 19.20      & 0.908      & -0.021     & 0.000       \\
 $ LR $&      & 19.24      & 0.899      & -0.023     & 0.001        \\
 $ LR1$&      & 19.26      & 0.899      & -0.019     & 0.000        \\
\end{tabular}
\end{table}

\begin{table}[]
\caption{CBF strenghts and positions of the one phonon DSF responses
with different correlations (see text). $q$ in $\AA^{-1}$ and
$\omega_q$ in $K$}
\begin{tabular}{lccccc}
        &$q$ & $\omega_q$ & $Z_{44}$ & $Z_{34}$ & $Z_{33}$ \\ \hline
 $ SR $&  1.1 & 13.73      & 0.275      & -0.066     & 0.016       \\
 $ LR $&      & 13.69      & 0.272      & -0.068     & 0.017       \\
 $ LR1$&      & 13.66      & 0.272      & -0.068     & 0.017       \\
 $ SR $&  1.3 & 14.01      & 0.390      & -0.045     & 0.005       \\
 $ LR $&      & 14.27      & 0.367      & -0.047     & 0.006       \\
 $ LR1$&      & 14.25      & 0.366      & -0.044     & 0.005       \\
 $ SR $&  1.5 & 13.84      & 0.559      & -0.014     & 0.000       \\
 $ LR $&      & 13.79      & 0.558      & -0.014     & 0.000       \\
 $ LR1$&      & 13.83      & 0.557      & -0.011     & 0.000       \\
 $ SR $&  1.7 & 13.94      & 0.766      &  0.024     & 0.001       \\
 $ LR $&      & 13.94      & 0.766      &  0.024     & 0.001        \\
 $ LR1$&      & 13.96      & 0.765      &  0.027     & 0.001        \\
\end{tabular}
\end{table}

\begin{table}[]
\caption{CBF sum rules $m_0$ and $m_1$ with the LR1 correlation and
CBF $^3$He spectrum, variational SSF's and $f$--sum rules.
$q$ in $\AA^{-1}$ and energies in $K$. In parentheses are shown the
multiphonon tail contributions}

\begin{tabular}{cccccc}
 $\alpha\beta$&$q$&$m_0(mPH)$&$m_1(mPH)$&$S_{\alpha\beta}$&$f-$SR \\\hline
   &1.1&      &            &      &       \\
  44 & & 0.37(0.08) & 7.38(3.58) & 0.36 & 7.29  \\
  34 & &-0.29(-0.08) &-4.80(-3.28) &-0.15 & 0.0   \\
  33 & & 1.10(0.13) & 8.88(4.67) & 0.98 & 9.73  \\
   &1.3&      &            &      &       \\
  44 & & 0.49(0.10) &10.27(4.97) & 0.46 &10.19  \\
  34 & &-0.24(-0.08) &-4.53(-3.22) &-0.12 & 0.0   \\
  33 & & 1.05(0.09) & 8.97(3.47) & 0.98 &13.59  \\
   &1.5&      &            &      &       \\
  44 & & 0.68(0.12) &13.59(5.86) & 0.68 &13.56  \\
  34 & &-0.14(-0.06) &-3.18(-2.53) &-0.07 & 0.0   \\
  33 & & 1.00(0.05) & 9.15(2.14) & 0.99 &18.09  \\
   &1.7&      &            &      &       \\
  44 & & 0.91(0.14) &17.40(6.72) & 0.90 &17.42  \\
  34 & &-0.04(-0.04) &-1.66(-1.86) &-0.02 & 0.0   \\
  33 & & 0.96(0.03) & 9.93(1.25) & 1.00 &23.24  \\

\end{tabular}
\end{table}


\begin{references}

\bibitem{Ebner70} C.Ebner and D.O.Edwards, Phys.Rep. C {\bf 2}, 77 (1970).

\bibitem{Baym78} G.Baym and C.Phetick in {\sl The Properties of Liquid
and Solid Helium}, Vol.2, ed.K.H.Benneman and J.B.Ketterson (Wiley, NY 1978).

\bibitem{DeBruyn87} R.De Bruyn Ouboter and C.N.Yang, Physica  B {\bf 144},
127 (1987).

\bibitem{Hilton77} P.A.Hilton, R.Scherm and W.G.Stirling,
J.Low.Temp.Phys. {\bf 27}, 851 (1978).

\bibitem{Bhatt78} R.N.Bhatt,
Phys.Rev. B {\bf 18}, 2108 (1978).

\bibitem{Fak90} B.F{\aa}k, K.Guckelsberger, M.Korfer, R.Scherm and A.J.Dianoux,
Phys.Rev. B {\bf 41}, 8732 (1990).

\bibitem{Sokol94} Y.Wang and P.E.Sokol, Phys.Rev.Lett. {\bf 72}, 1040 (1994).

\bibitem{Szprynger85} A.Szprynger and M.Lucke, Phys.Rev. B {\bf 32}, 4442
(1985).

\bibitem{Saarela_rep} E.Krotscheck and M.Saarela, Phys.Rep.  {\bf 232}, 1
(1993).


\bibitem{Hsu} W.Hsu, D.Pines and C.H.Aldrich, Phys.Rev. B {\bf 32}, 7179
(1985).

\bibitem{FantoniHe4} Q.N.Usmani, S. Fantoni and V.R. Pandharipande,
Phys.Rev. B {\bf 26}, 6123 (1982).

\bibitem{FantoniHe3} E. Manousakis, S. Fantoni, V.R. Pandharipande, and Q.N.
Usmani, Phys.Rev. B {\bf 28}, 3770 (1983).

\bibitem{VivianiHe3} M. Viviani, E. Buend\'\i a, S. Fantoni, and S. Rosati,
Phys.Rev. B {\bf 38}, 4523 (1988).

\bibitem{Fabrocini86} A. Fabrocini, S. Fantoni, S. Rosati, and A. Polls,
Phys.Rev. B {\bf 33}, 6057 (1986).

\bibitem{Greywall79} D.S.Greywall, Phys.Rev. B {\bf 20}, 2643 (1979).

\bibitem{Owers88} J.R.Owers-Bradley, P.C.Main. R.M.Browley, G.J.Batey
and R.J.Church, J.Low Temp.Phys. {\bf 72}, 201 (1988).

\bibitem{Landau48} L.D.Landau and I.M.Khalatnikov, Zh.Eksp.Teor.Fiz.
 {\bf 19}, 637 (1948).

\bibitem{Feenberg} E.Feenberg, {\sl Theory of Quantum Fluids}
(Academic Press, NY 1969).

\bibitem{FabrociniRL} A.Fabrocini and S.Fantoni,
  Nucl. Phys. {\bf A503}, 375 (1989).

\bibitem{R_spin} A.Fabrocini, Phys.Lett. {\bf B322}, 171 (1994).

\bibitem{RL_nuclei} O.Benhar, A.Fabrocini, S.Fantoni and I.Sick,
  Nucl. Phys. {\bf A579}, 493 (1994).

\bibitem{Manousakis86}  E.Manousakis and V.R.pandharipande,
Phys. Rev. {\bf B33}, 150 (1986).

\bibitem{Sears86} V.F.Sears in {\sl Neutron Scattering},
 Vol.23A of {\sl Methods of Experimental Physics}, ed.K.Skold and
 D.L.Price (Academic, NY 1986).

\bibitem{Feynman} R. P. Feynman, Phys. Rev. {\bf 94}, 262 (1954)

\bibitem{Boronat93} J. Boronat, A. Polls and A. Fabrocini,
J. Low Temp. Phys. {\bf 91}, 275 (1993).

\bibitem{Owen81} J.C.Owen ,
Phys. Rev. {\bf B23}, 5815 (1981).

\bibitem{FP84} A. Fabrocini and  A. Polls ,
Phys. Rev. {\bf B30}, 1200 (1984).

\bibitem{Chakraborty83} T.Chakraborty, A.Kallio, L.J.Lannto and P.Pietilainen ,
Phys. Rev. {\bf B27}, 3061 (1983).

\bibitem{Lee75} D.K.Lee and F.J.Lee, Phys. Rev. {\bf B11}, 4318 (1975).

\bibitem{VRP_eps} B.Friedman and V.R.Pandharipande,
Phys. Lett. {\bf B100}, 205 (1981).

\bibitem{Fabrocini_prep} A.Fabrocini, L.Vichi, F.Mazzanti and A.Polls, in
preparation.

\bibitem{Fak_priv}  B.F{\aa}k, private communication.



\end{references}
\end{document}